\documentclass[a4paper,12pt]{article}

\makeatletter
\renewcommand\paragraph{\@startsection{paragraph}{4}{\z@}%
  {-3.25ex\@plus -1ex \@minus -.2ex}%
  {1.5ex \@plus .2ex}%
  {\normalfont\normalsize\bfseries}}
\makeatother

\makeatletter
\renewcommand\subparagraph{\@startsection{subparagraph}{4}{\z@}%
  {-3.25ex\@plus -1ex \@minus -.2ex}%
  {1.5ex \@plus .2ex}%
  {\normalfont\normalsize\bfseries}}
\makeatother

\usepackage[round, sort&compress]{natbib}
\usepackage[latin1]{inputenc}
\usepackage[english]{babel}
\usepackage{theorem}
\usepackage{amsmath}
\usepackage{amsfonts}
\usepackage{amssymb}
\usepackage{amscd}
\usepackage{amstext}
\usepackage{amsbsy}
\usepackage{amsopn}
\usepackage{amsxtra}
\usepackage{color}
\usepackage{euscript}
\usepackage{empheq}
\usepackage{upref}
\usepackage{graphicx}
\usepackage{subfig}
\usepackage{verbatim}
\usepackage[francais,verbose]{layout}
\usepackage{fancyhdr}
\usepackage{psfrag}
\usepackage{multirow}
\usepackage{authblk}

\theorembodyfont{ \upshape}% écrit le texte des exemples, des exercices et des preuves en plaintext

\theoremstyle{break} % écrit à la ligne dans les exemples et exercices

\newtoks\ladate
\newtoks\sujet

pt\belowdisplayshortskip 5 pt plus 3 pt minus 3 pt

\begin{document}

\title{When games meet reality: is Zynga overvalued?}
\author{Zal\'an Forr\'o\thanks{zforro@ethz.ch, +41 44 632 09 28}}
\author{Peter Cauwels\thanks{pcauwels@ethz.ch, +41 44 632 27 43}}
\author{Didier Sornette\thanks{dsornette@ethz.ch, +41 44 632 89 17}}
\affil{Department of Management, Technology and Economics, ETH Z\"urich}
%\author{Zal\'an Forr\'o, Peter Cauwels and Didier Sornette}
%\email{zforro@ethz.ch, pcauwels@ethz.ch, dsornette@ethz.ch}
\maketitle

\newpage

\begin{abstract}
\noindent On December 16\textsuperscript{th}, 2011, Zynga, the well-known social game developing company went public. This event followed other recent IPOs in the world of social networking companies, such as Groupon or Linkedin among others. With a valuation close to 7 billion USD at the time when it went public, Zynga became one of the biggest web IPOs since Google. This recent enthusiasm for social networking companies raises the question whether they are overvalued. Indeed, during the few months since its IPO, Zynga showed significant variability, its market capitalization going from 5.6 to 10.2 billion USD, hinting at a possible irrational behavior from the market. To bring substance to the debate, we propose a two-tiered approach to compute the intrinsic value of Zynga. First, we introduce a new model to forecast its user base, based on the individual dynamics of its major games. Next, we model the revenues per user using a logistic function, a standard model for growth in competition. This allows us to bracket the valuation of Zynga using three different scenarios: 3.4, 4.0 and 4.8 billion USD in the base case, high growth and extreme growth scenario respectively. This suggests that Zynga has been overpriced ever since its IPO. Finally, we propose
an investment strategy (dated April 19\textsuperscript{th}, 2012 on the arXive), which is based on our diagnostic of a bubble for Zynga  and how this herding / bubbly sentiment can be expected to play together with
two important coming events (the quarterly financial result announcement around April 26\textsuperscript{th}, 2012
followed by the end of a first lock-up period around April 30\textsuperscript{th}, 2012). On the long term,
our analysis indicates that Zynga's price should decrease significantly. 
The paper ends with a post-mortem analysis added on May 24\textsuperscript{th}, 2012, 
just before going to press, showing that
we have successfully predicted the downward trend of Zynga. Since April 27\textsuperscript{th}, 2012, Zynga dropped 25\%.
\end{abstract}


\newpage
\section{Introduction}
\noindent After the recent initial public offerings (IPOs) of some of the major social networking companies such as Groupon, Linkedin or Pandora, Zynga went public on December 16\textsuperscript{th}, 2011. In November 2011, the estimated value of this social network game developing company was as high as 14 billion USD \citep{YahWeb2}. However, after the underperformance of the IPO market, this number was scaled down. Indeed, 100 million shares of Class A common stock were sold at $10\$$ per share, the top end of the indicative $8.5\$$ -- $10\$$ range (\citet{SecWeb}, 2011). With a total of 699 million shares outstanding, the market capitalization of the company at IPO was of 7 billion USD. After dropping to 5.6 billion USD on January 9\textsuperscript{th}, 2012, its minimum since the IPO, and peeking at 10.2 billion USD on February 14\textsuperscript{th}, 2012, the market capitalization of Zynga was around 9 billion USD on February 26\textsuperscript{th} \citep{BloWeb}. The efficient-market hypothesis suggests that such large price changes should reflect significant variations in the fundamentals of the company that lead to the re-assessment of its value by investors and analysts \citep{Fam70}. As such, one could question the economic justification for such a change in price in only a few months time. \\

In addition, during the IPO process, no specifics were given on the methodology to come to Zynga's valuation; neither by the underwriters of the IPO, nor in the S1 filing, nor by the media, nor by investment banking sell-side analysts. The aim of this paper is to determine Zynga's fundamental value and put its current valuation into perspective. For this purpose, we extend to Zynga the methodology proposed by \cite{CauSor12} for the valuation of Facebook and Groupon, by introducing a semi-bootstrap approach to forecast Zynga's user base. \\ 

The pricing of IPOs, and companies in general, has been extensively studied. \cite{IbbRit95,RitWel02}, among others, reviewed well-known stylized facts when companies go public. We can cite the underpricing of new issues, ie, the fact that underwriters underprice the IPO leading to high returns on the first day of trading, or the long-term underperformance of the underpriced IPOs compared to their ``fairly" priced counterparts. During the dot-com bubble, the rapid rise of the Internet sector contrasting with the modest growth of the ``old economy" raised a lot of interest. \cite{Baretal02} showed that there were differences in the valuation of Internet and non-Internet firms. Notably, for the latter, profits were rewarded (positively correlated with the share value) and losses were not (as is usually the case). However, the reverse was true for Internet companies, where losses were rewarded and profits were not. This somewhat paradoxical situation arose from the perception that losses were not the result of poor company management but rather investments that would later pay off. \cite{Han01, DemLev00} further showed that web-traffic was an important factor in the market value of the Internet company. Indeed, in the case of web-traffic intensive companies, while losses were being rewarded before the peak of the bubble and profits were not, the situation reversed after the peak of the bubble: profits became rewarded and losses not anymore. This phenomenon was not observed for Internet companies without web-traffic. \\

We should notice that the studies, mentioned so far, tried to explain the market price of companies using different explanatory variables (such as revenues, type of company, amount of web-traffic, difference between IPO price and first day closing price and so on), making the implicit assumption that the market is efficient and reflects the intrinsic value of the company. While in the long-run this may be a good approximation, this is not true for shorter time-scales (during a bubble typically). As such, these methods (often based on linear relationships between the market price and the explanatory variables) are not meant to reveal the fundamental value of a company or make long-term predictions. \\ %but rather to inform us about the relevant factors influencing the value of a company in the eyes of the market during a given period.

 \cite{OfeRic02} tackled the problem from a different angle. They assumed that, on the long-run, the price-to-earnings ratio of the Internet companies would converge to their ``old economy" counterparts, and computed the growth in earnings necessary to achieve that. They found unrealistic growth rates making an argument against market rationality. \cite{SchMoo00} used a real-option approach to value Amazon, the company having the option to go bankrupt (thus limiting their losses). Their model relies on the future growth rate of revenues and use the discounted cash flows method. It has the upside to come up with a valuation for the company but is very sensitive to variations in its parameters. \cite{Gupetal04} extended a methodology developed by \cite{Kimetal95} to value Amazon, Ameritrade, E-bay and E*Trade (Internet companies). Their model uses  the discounted cash flows analysis where the future revenues are computed based on the prediction of the company's user base combined with an estimation of the revenues generated by each user in time. They obtain robust valuations of the Internet companies, allowing for a quantitative assessment of the discrepancy between the market capitalization of the companies and their fundamental values. Adopting a similar approach, \cite{CauSor12} show that Facebook and Groupon are overvalued. The main insight of the aforementioned works is to recognize that, for companies deriving their value directly from their users, such a simple approach can give much better estimate of the intrinsic value of a company than methods employed so far. \\
 
The present paper adds to the existing literature by extending the methodology of \cite{CauSor12} to Zynga, a company where the user dynamics are very different and more complicated from the ones observed in Facebook, Groupon, Amazon and so on. Indeed, the evolution of Zynga's user base is a result of the individual dynamics of each of its individual games and cannot therefore be modeled by a single function. Moreover, we find that the revenues per user have entered a saturation phase. This limits Zynga's ability to increase its revenues much further, as their user base is already in a quasi-stationary phase. Finally, we find that Zynga has been greatly overvalued since its IPO and give a short time scale prediction about its price dynamics by combining our fundamental analysis with the effect consisting of a price drop subsequent to the end of the lock-up period.\\

\noindent This paper is organized as follows. Section \ref{methodology} gives a brief summary of the methodology used to value Zynga. Section \ref{users} describes the dynamics of the number of daily active users (DAU) of Zynga. Section \ref{financials} analyzes the financial data relevant to the valuation of the company. Section \ref{valuation} gives its estimated market capitalization. Section \ref{comparison} analyzes the evolution of Zynga so far in the light of its valuation.
Section \ref{arbitraging} discusses possible strategies to arbitrage the
over-valued stock of Zynga and section \ref{conclusion} concludes.

\section{ Valuation methodology} \label{methodology}
\begin{figure}[!h] 
    \centering
    \psfrag{1}[B][B][1.2][0]{s(t)}
    \includegraphics[width=1\textwidth]{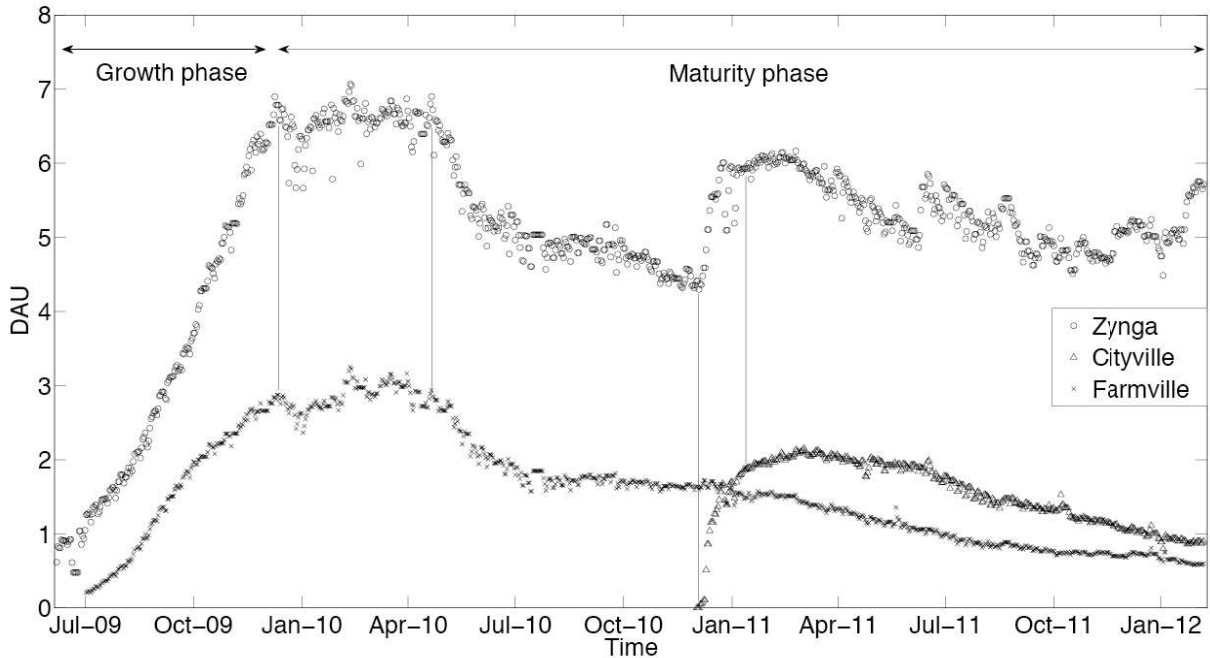}
    \caption{The number of DAU as a function of time for Zynga and two of its most popular games, Cityville and Farmville. After an initial growth period, Zynga entered a quasi-stable maturity phase since January 2010. A typical feature of the games can be seen in Cityville and Farmville: after an initial rapid rise, the DAU of the games enters a slower decay phase. The black vertical lines show that the total DAU of Zynga depends stronly on the performance of the underlying games. Notice that, even though Zynga exists since mid-2007, we do not have DAU data since its beginning. (Source of the data: \emph{http://www.appdata.com/devs/10-zynga})}
    \label{dau_time}
\end{figure}

The major part of the revenues of a social networking company is inherently linked to its user base. The more users it has, the more income it can generate through advertising. From this premise, the basic idea of the method proposed by \cite{CauSor12} is to separate the problem into 3 parts:
\begin{enumerate}
\item First, we will forecast Zynga's user base. This is what we call the part of the analysis based on hard data and modeling. Because Zynga uses Facebook as a platform, and one does not need to register to have access to its games (a facebook account is sufficient), there is no such thing as a measure of total registered users (U). These registered users were used by \cite{CauSor12} in their valuation of Facebook and Groupon. Because it takes an effort to unregister, the number of registered users is an almost monotonically increasing quantity. As such, \cite{CauSor12} were able to model and forecast Facebook and Groupon's user dynamics with a logistic growth model (equation \ref{logistic}).

\begin{equation} \label{logistic}
\frac{dU}{dt} = gU(1 - \frac{U}{K})
\end{equation}

\noindent Here, $g$ is the constant growth rate and $K$ is the carrying capacity (this is the biggest possible number of users). This is a standard model for growth in competition. When $U \ll K$, U grows exponentially since $\frac{dU}{dt} \approx gU$ (this is the unlimited growth paradigm) until reaching saturation when $U = K$ (and $\frac{dU}{dt} = 0$). This model is a good description of what happens in most social networks: the number of users starts growing exponentially and eventually saturates because of competition/constrained environment.\\
For Zynga, a different approach had to be worked out. Here, the analysis is based on the number of Daily Active Users (DAU), a more dynamical measure. DAU can fluctuate (up and down) and as such cannot be modeled with a logistic function. Moreover, Zynga's users form an aggregate of over 60 different games. Therefore, to understand the dynamics of Zynga's user base, we had to examine the user dynamics of its individual games. Figure \ref{dau_time} gives the total number of Zynga users and the DAU of two of its most popular games. We decided to model each of its top 20 games individually, this approach accounting for more than 98$\%$ of the recent total number of Zynga users. The specifics of this analysis will be further elaborated in section \ref{users}.   
 
\item The second part of the methodology is based on what we consider as ``soft" data: this part uses the financial data available in the S1/A Filing to the SEC (2011). These will be used to estimate the revenues that are generated per daily active user in a certain time period. It also reveals information on the profitability of the company. Due to the limited amount of published financial information, we will have to rely on our intuition and good-sense to give our best estimate of the future revenues per user generated by the company. This is why we call this part the soft data part. It will be further elaborated in section \ref{financials}.

\item The third part combines the two previous parts to value the company. With an estimate of the future daily number of users (DAU) and of the revenues each of them will generate $(r)$, it is possible to compute the future revenues of the company. These are converted into profits using a best-estimate profit margin ($p_{\text{margin}}$), and are discounted using an appropriate risk-adjusted return $d$. The net present value of the company is then the sum of the discounted future profits (or cash flows).

\begin{equation} \label{eq:valuation}
Valuation = \sum_{t=1}^{end}  \frac{r(t) \cdot DAU(t) \cdot p_{\text{margin}}}{(1+d)^t} = \sum_{t=1}^{end} \frac{\text{profits}(t)}{(1+d)^t}
\end{equation}

\noindent Hereby, we optimistically assume that all profits are distributed to the shareholders.
\end{enumerate}

\section{Hard Data} \label{users}

\subsection{General approach}
We will take the following steps to forecast Zynga's DAU: 
\begin{enumerate}
\item We will use a functional form to the DAU of each of the top 20 games to forecast the future DAU evolution of the company. This is done as follows:
	\begin{itemize}
	\item The data that are available are used as it is.
	\item This is extended into the future by extrapolating the DAU-decay process with an appropriate tail function.
	\end{itemize}
\item Because Zynga relies on the creation of new games in order to maintain or even increase its user base, it is important to quantify its rate of innovation. This is done by using $p(\Delta t)$ the probability distribution of the time between the implementation of 2 consecutive new games (restricted to the top 20).
\item Finally, a future scenario is simulated as follows: for the next 20 years, each $\Delta t$ days, $\Delta t$ being a random variable taken from $p(\Delta t)\text{,}$ a game is randomly chosen from our pool of top 20 games. The DAU of Zynga over time is then simply the sum of the simulated games. A thousand different scenarios are computed.
\end{enumerate}

\subsection{The tails of the DAU decay process}
\begin{figure}[!h]
    \centering
    \includegraphics[width=1 \textwidth]{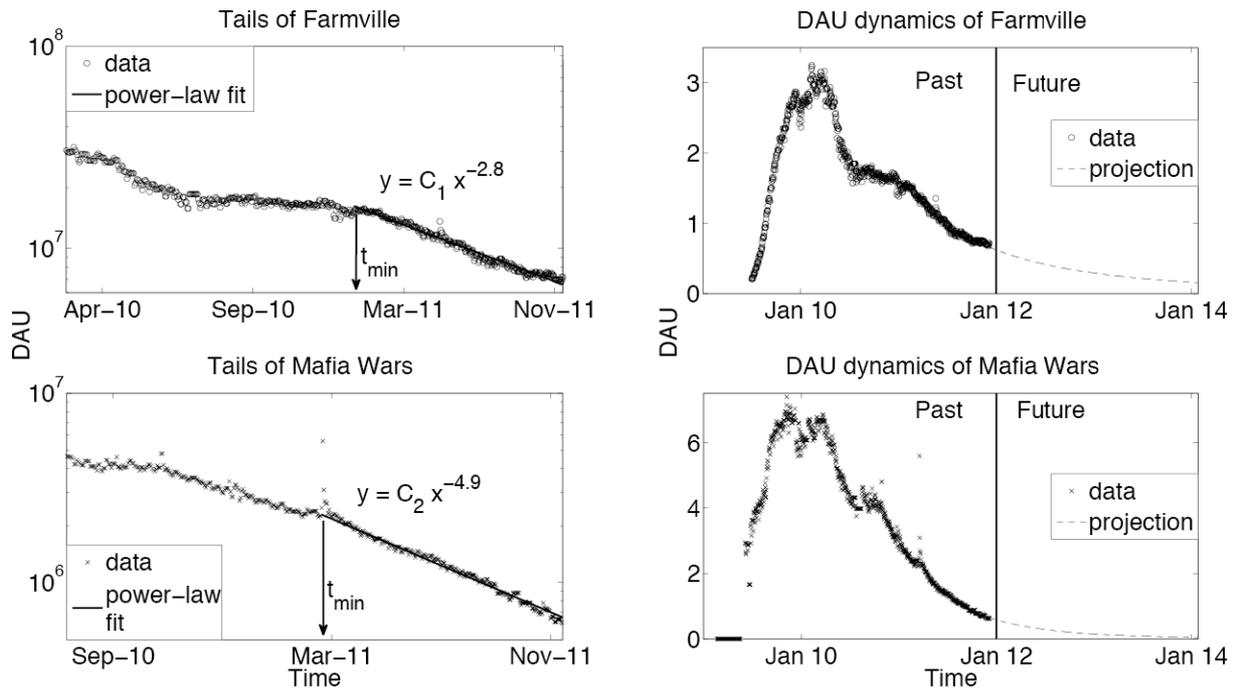}
    \caption{Left: Decay of Farmville (circles) and Mafia Wars (crosses), 2 representative games out of Zynga's top 20. It can be seen that a power law is a good fit for the tails from $t_{min}$ onwards. Right: Simulated dynamics of the games based on the power law parameters of the left panel. (Source of the data: \emph{http://www.appdata.com/devs/10-zynga})}
    \label{tails}
\end{figure}

The functional form of the DAU of each game is composed of the actual observed data and a tail that simulates the future decay process. We will use a power law, $f(t) \propto t^{-\gamma}$, for that purpose. This results in a slow decay process and as such will not give rise to any unnecessary devaluation of the company by underestimating its future user base. Figure \ref{tails} shows the power law fits (left) and the extension of the user dynamics into the future (right) for the games Farmville and Mafia Wars.

Such power law is a reasonable prior, given the large evidence of such time dependence in many human activities \citep{sornette2005origins},
which includes the rate of book sales \citep{sornette2004amazon,deschatres2005amazon}, the dynamics of 
video views on YouTube  \citep{cranesorYouTube}, the dynamics of visitations of major news portal \citep{Dezso_et_al_06}, 
the decay of popularity of Internet blogs posts \citep{Leskovec_et_al_07}, the rate of donations
following the tsunami that occurred on December 26, 2004
\citep{Crane_DonationResponse} and so on.

\subsection{Innovating process}
To be able to realistically simulate Zynga's rate of innovation, it is important to understand the generating process underlying the creation of new games. The simplest process that can be used for that purpose is the Poisson process. To understand its meaning, consider the Bernouilli process, its discrete counterpart. It has a very intuitive meaning and can be thought of as follows: at each time step, a game is introduced with a probability of $p$ (and no game is introduced with a probability of $1-p$). For a large enough number of time steps and a small enough $p$, the Bernoulli process converges to the Poisson process. The Poisson process has 3 important properties:
\begin{enumerate}
\item It has a constant innovation rate. 
\item It has independent inter-event durations. 
\item The inter-event durations have an exponential distribution: $p(\Delta t) = e^{-\lambda \Delta t}$.
\end{enumerate}

\noindent To assess whether this is a suitable process to model the innovation rate, we measured the time between the introduction of two consecutive new games, $\Delta t_{(1,2)}, \Delta t_{(2,3)}, ... ,\Delta t_{(n-1, n)}$, and tested for the above mentioned 3 properties.

\subsubsection{Innovation rate}
\begin{figure}[!h] 
    \centering
    \includegraphics[width=1\textwidth]{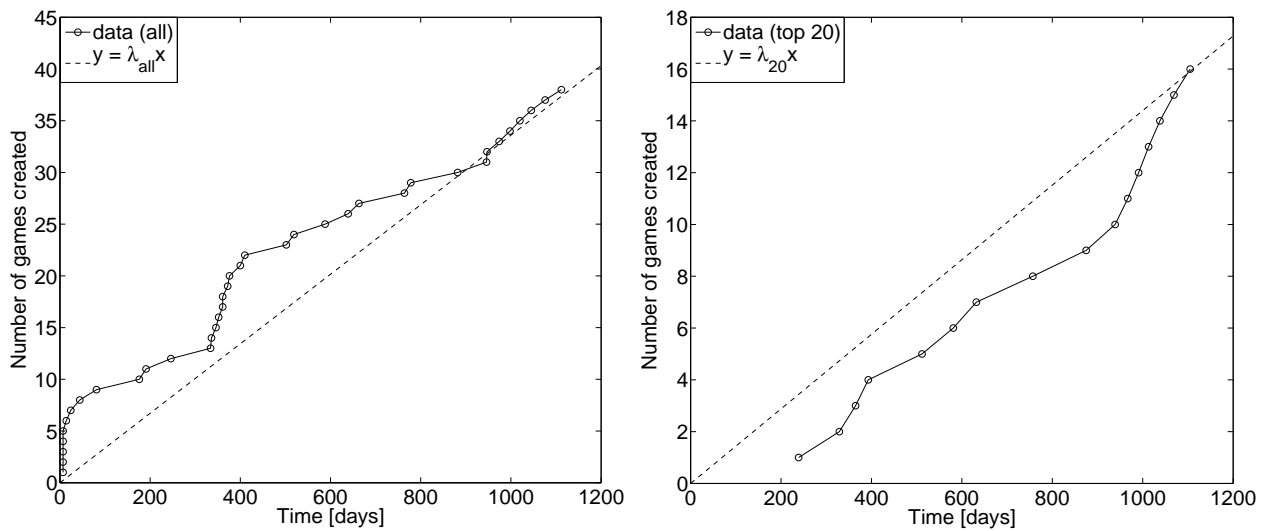}
    \caption{Left: number of newly created games as a function of time for all the games. The empirical innovation rate is at most equal to the theoretical rate coming from the Poisson process (dashed line). The parameter $\lambda_{all}$ is obtained by using maximum likelihood (assuming a Poisson process). Right: number of newly created games as a function of time for the top 20. The empirical innovation rate seems to be higher for the last games than $\lambda_{20}$, the theoretical rate from the Poisson process. This is most likely due to insufficient statistics, this phenomenon being absent when all games are taken into account.}
    \label{counting}
\end{figure}

To test whether the innovation rate is constant, different approaches can be adopted. One possibility is to test for the stationarity of the DAU of Zynga since it entered its maturation phase. Stationarity in the number of users would imply a constant innovation rate. Indeed, figure \ref{dau_time} suggests that the user dynamics of Zynga are stationary, its number of DAU being comprised between 43 and 70 million users since the end of the growth phase. However, due to the short time-span of the data, it is hard to implement rigorous statistical tests such as unit-root tests. Instead, we adopt a different approach. If the rate of creation of new games is constant, then the number of new games created as a function of time should lie around a straight line with slope $\lambda$, the intensity of the Poisson process. Figure \ref{counting} shows the counting of new games as a function of time.

\noindent As we can see from figure \ref{counting}, the constant innovation rate is a good approximation. Our main concern was to discard the possibility of an important increase in the frequency of creation of new games towards the end of the time period, which would have lead to an underestimation of the number of new games created in the future and hence of the future number of users. When all games are taken into account, the innovation rate is at most equal to the one coming from the Poisson process. As such, the Poisson process with constant intensity would not lead to an underestimation of the value of the company. 

\subsubsection{Independence of inter-event times} \label{independence}
\begin{figure}[!h] 
    \centering
    \includegraphics[width=1\textwidth]{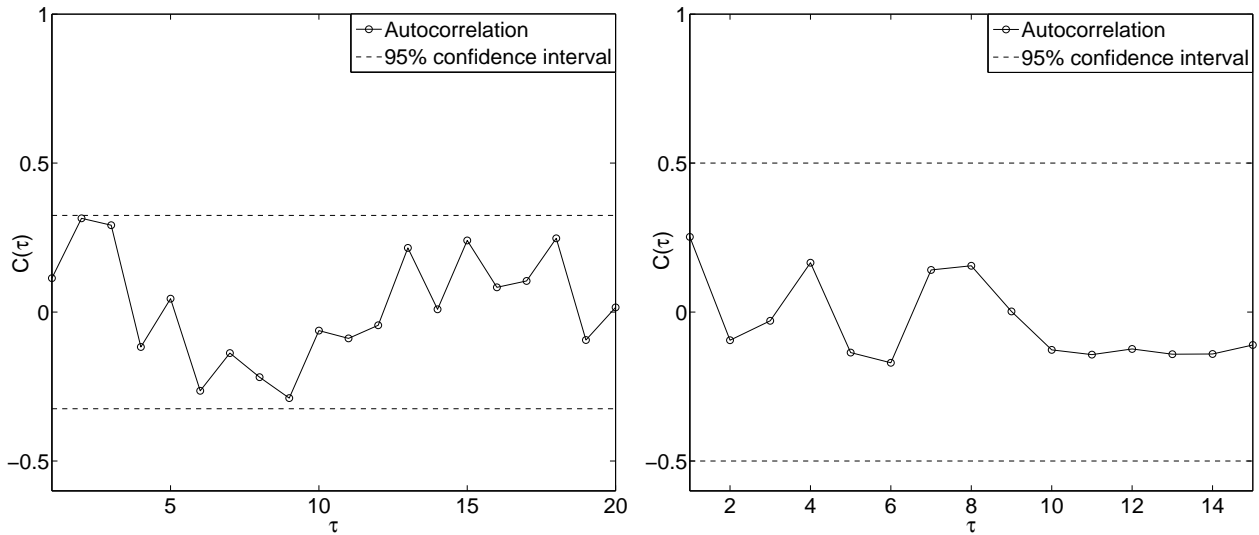}
    \caption{Left: Autocorrelation function for the inter-event times of all the games. The confidence interval (CI) indicates the critical correlation needed to reject the hypothesis that the inter-event times are independent. It is computed as $CI_{.95} = \pm \frac{2}{\sqrt{N}}$, $N$ being the sample size \citep{Cha04}. Right: Autocorrelation function for the inter-event time of the top 20 games. In both cases, the independence hypothesis cannot be rejected.}
    \label{corr_20_and_all}
\end{figure}

To test for the independence of the measured $\Delta t$, we study the autocorrelation function $C(\tau)$, the correlation between $\Delta t_{(i,i+1)}$ and $\Delta t_{(i+\tau,i+\tau+1)}$. The result is given in figure \ref{corr_20_and_all}.

\noindent As can be seen, $C(\tau)=0$ is within the confidence interval for $\tau > 0$ in both cases, so the independence hypothesis cannot be rejected.
\subsubsection{Distribution of inter-event times}
\begin{figure}[!h] 
    \centering
    \includegraphics[width=1\textwidth]{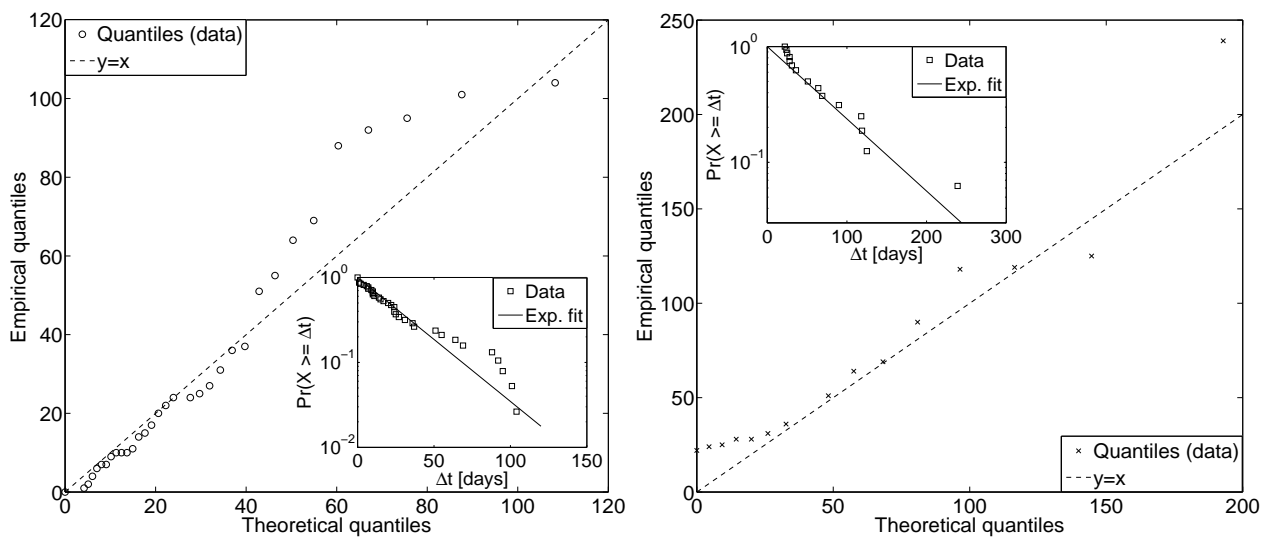}
    \caption{Left: Q-Q plot of the distribution of time intervals $\Delta t$ between the introduction of new games, for all the games. The theoretical (assuming a Poisson process) and data quantiles (red circles) agree well as can be seen from the proximity to the dashed black line (y=x). The subpanel shows the CDF of the data (red squares) and the exponential fit (black line) on which the quantiles were built. Right: Q-Q plot of $\Delta t$ for the top 20 games. We can see a deviation for small values of $\Delta t$ between exponential theoretical and data quantiles. This can be attributed to insufficient statistics.}
    \label{qq_20_and_all}
\end{figure}

To test for the distribution of inter-event times, we use a Q-Q plot. The Q-Q plot is a graphical method for comparing two distributions  by plotting their quantiles against each other. If the obtained pattern lies on a straight line, the distributions are equal. In our case, the two distributions to be compared are the empirical one (from data) and the theoretical one, an exponential with parameters obtained from a maximum likelihood fit to the data. The results of this analysis are presented in figure \ref{qq_20_and_all}.
 
\noindent As we can see, in both cases the Q-Q plots show a reasonable agreement between the empirical and the exponential distribution given the number of data points, so that the exponential distribution for the inter-event times $(\Delta t)$ cannot be rejected. The innovation process will thus be modeled as a Poisson process.

 \subsection{Predicting the future DAU of Zynga}
\begin{figure}[!h] 
    \centering
    \includegraphics[width=1\textwidth]{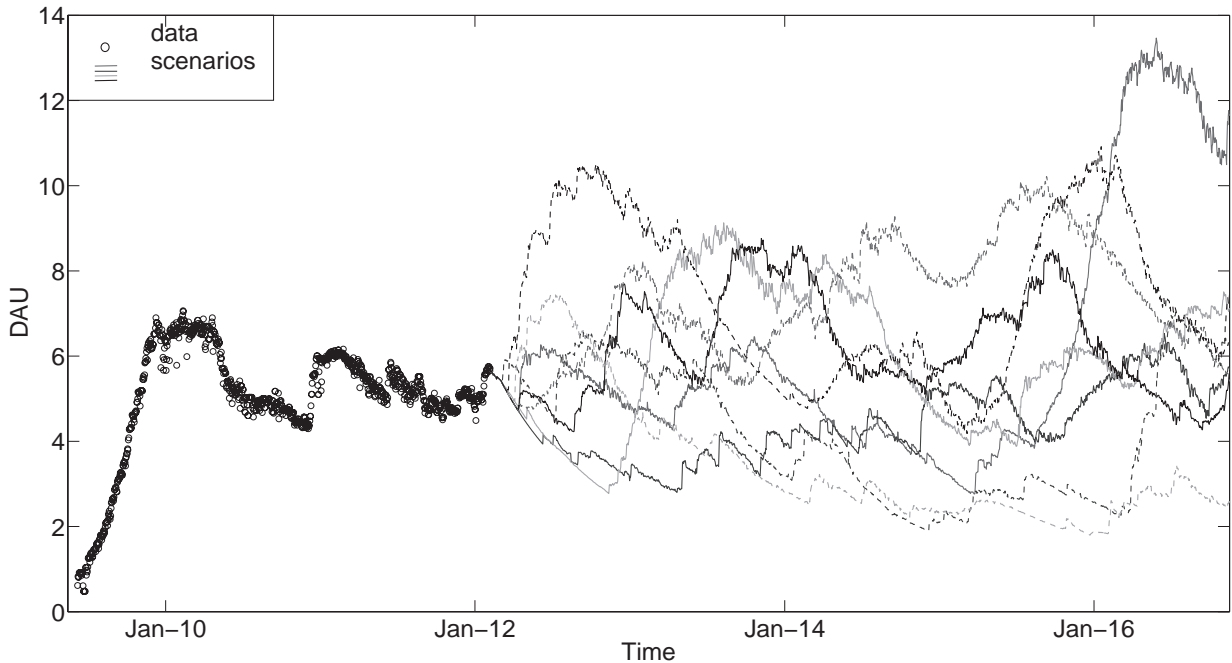}
    \caption{8 different scenarios of the DAU evolution of Zynga for the next 4 years. The company will be valued for each of these scenarios (section \ref{valuation}). This will give a range for its expected market capitalization.}
    \label{scenarios}
\end{figure}

Starting from the present up to the next 20 years, a top 20 game is randomly sampled each $\Delta t$ days, with $\Delta t$ drawn from its theoretical exponential distribution $p(\Delta t)$. For each of these sampled games, the DAU is calculated using its functional form. Summing the DAU of all these games, the user's dynamics of Zynga is computed. This process is repeated a thousand times, giving a thousand different scenarios. As can be seen from figure \ref{scenarios}, the evolution of the user base between scenarios can be quite different. That is the reason why a wide range of scenarios are needed. The valuation of the company will be computed for each of those scenarios (using equation \ref{eq:valuation}). This will give a probabilistic forecast of the market capitalization of Zynga. 

\section{Soft Data} \label{financials}
The next step to calculate Zynga's value is to estimate the revenues per DAU per year. We base our analysis on the S1/A Form (2011) complemented with the 8-K Form of Filings to the SEC (2012), to add the last quarter of 2011 results. The yearly revenues are given each quarter as a running sum of the four previous quarterly revenues:
\begin{equation}
R_i = R^q_{i-3} + R^q_{i-2} + R^q_{i-1} + R^q_{i}~.
\end{equation}

\begin{figure}[!h] 
    \centering
    \includegraphics[width=1\textwidth]{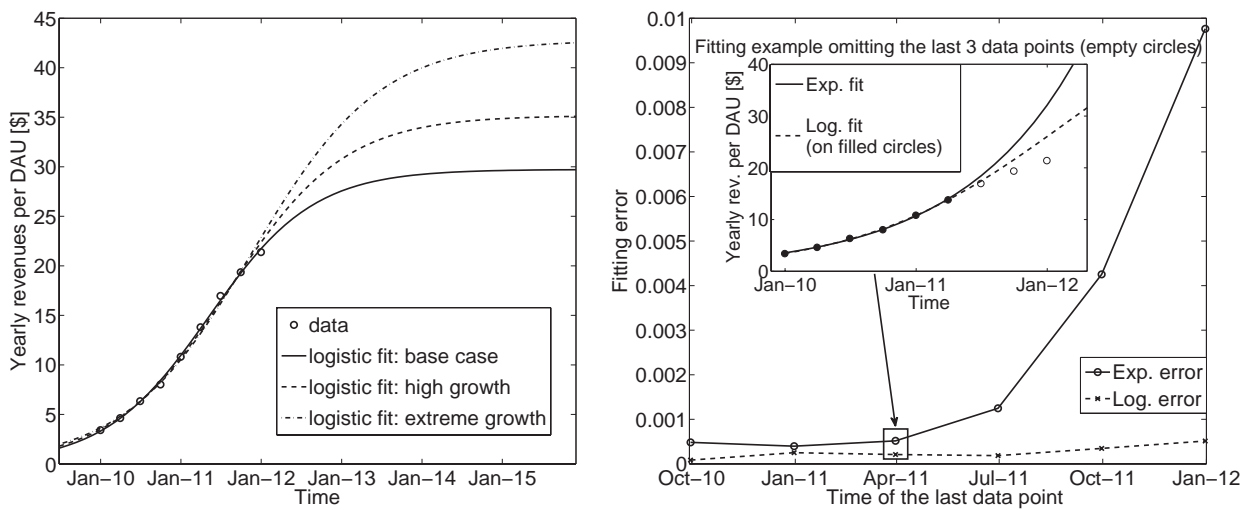}
    \caption{Left: Yearly revenue per DAU over time. A logistic fit (equation \ref{logistic}) is proposed with $K \approx 30, 35$ and $43$USD for the base case, high growth and extreme growth scenarios. Right: Fitting error of the exponential vs logistic function. Each point is obtained by performing the logistic and exponential regressions on data taking more and more data points into account, starting from a minimum of 4 (as shown in the subpanel where only filled circles are fitted). We can see that the logistic starts performing significantly better than the exponential from July 2011. (Source of the data: S1/A and \citet{SecWeb2}.)}
    \label{rev_per_dau}
\end{figure}

\noindent Here, $R_i$ and $R^q_i$ are respectively the yearly and quarterly revenues at quarter $i$ with $i \in (4, \text{last})$. The yearly revenues per DAU at each quarter, $r_i$, are then obtained by dividing $R_i$ by $\langle DAU_i \rangle_{year}$, the realized DAU at time $i$ averaged over the preceding year. Figure \ref{rev_per_dau} gives the historical evolution of the revenues per DAU. Initially, this followed an exponential growth process. However, as can be clearly seen in the right panel, this growth is saturating and the process is following the trajectory of a logistic function. This implies that the revenues per DAU will reach a ceiling. As such, different logistic functions are fit to the dataset. Each of these corresponds to a different scenario: a base case, a high growth and an extreme growth scenario (as defined in \citealp*{CauSor12}). They can be seen on the left panel. \\

\noindent From the beginning of Zynga up to April of 2011, both the exponential and the logistic fits perform similarly. Indeed, when $r_i \ll K$, when the revenues per user are far away from saturation, the logistic function can be approximated by an exponential. April 2011 is a turning point in the sense that the growth of the revenues per user slows down, hence the deviation from the exponential (growth at constant rate). This saturation in $r_i$ is easy to explain: there have to be constraints on how much money can be extracted from a user. Under spatial constraints (there is a limited number of advertisements that can be displayed on a webpage), time constraints (there are only so many advertisements that can be shown per day) and ultimately the economic constraints (there is only so much money a user can spend on games or an advertiser is willing to spend), the revenues per DAU are bound to saturate. Using this logistic description for the revenues per DAU, a valuation of Zynga will be given for each of the 3 growth hypotheses.

\section{Valuation} \label{valuation}
Combining,  through equation \ref{eq:valuation}, the hard part of the analysis with the soft part, ie, the number of users over time and the revenues each of them generates per year, the value of the company can be calculated.

\begin{figure}[!h] 
    \centering
    \includegraphics[width=1\textwidth]{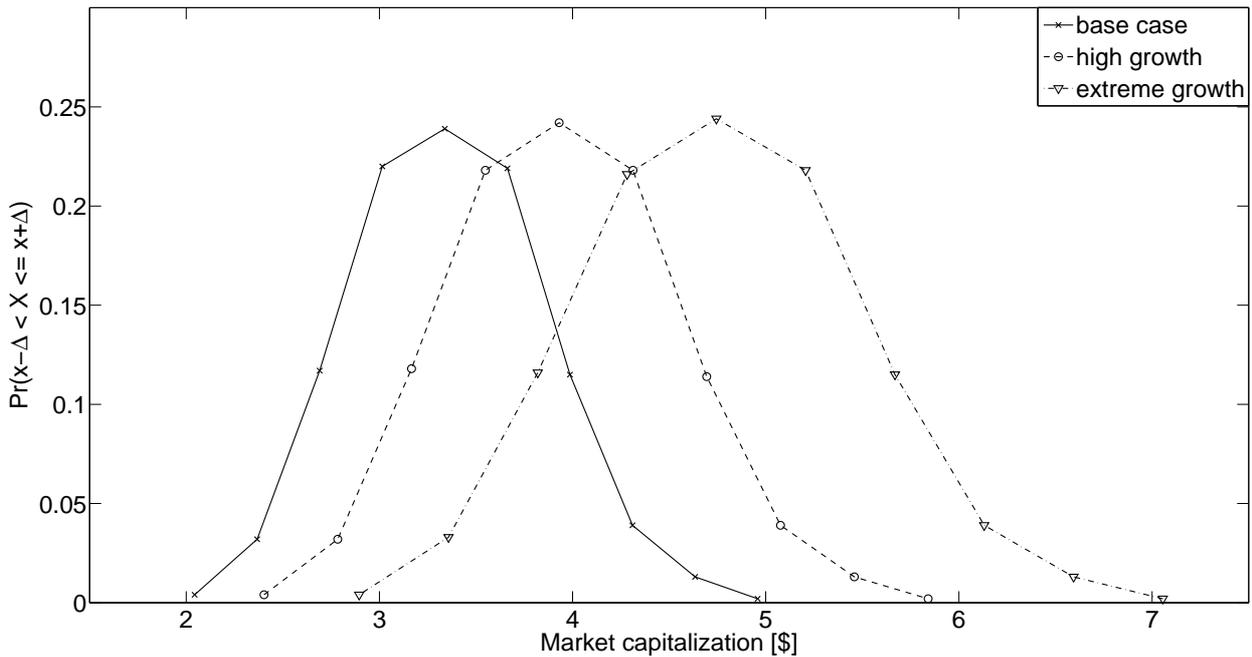}
    \caption{Distribution of the market capitalization of Zynga according to the base case, high growth and extreme growth scenarios. This shows that the 7 billion USD valuation at IPO or today's 9 billion valuation (March 2012) is not even satisfied in the extreme revenues case.}
    \label{fig:valuation}
\end{figure}

\noindent We will use a profit margin of $15\%$. This is Zynga's profit margin of fiscal year 2010. As can be seen from table \ref{profit_margin}, this is an optimistic assumption since it was the highest profit margin until now, 2010 being the only profitable year of Zynga so far. We also assume that all profits will be distributed to the shareholders and use a discount factor of $5 \%$ as in \cite{CauSor12}. We computed the company's valuation for all 1000 different scenarios using equation \ref{eq:valuation}. The results are shown in figure \ref{fig:valuation} and table \ref{table}.

\begin{table} [!h]
    \centering
    \begin{tabular}{ | l | c | c | c | c |}
    \hline
    Year & 2008 & 2009 & 2010 & 2011 \\ \hline
    Revenue (millions USD) & 19.41 & 121.47 & 597.46 & 1065.65 \\ \hline
    Net income (millions USD) & -22.12 & -52.82 & 90.60 & -404.32 \\ \hline
    Profit margin& $-114\%$ & $-43\%$ & $15\%$ &  $-38\%$  \\ \hline
    \end{tabular}
    \caption{Revenue, net income and profit margin of Zynga. (Source of data: S1/A and 8-K forms of the filings to the SEC.)}
    \label{profit_margin}
\end{table}

\begin{table} [!h]
    \centering
    \begin{tabular}{ | l | c | c | c | c |}
    \hline
    Scenario & Valuation [$\$$] & $95\%$ conf. interval & Share [$\$$] & $95\%$ conf. interval \\ \hline
    Base case & 3.4 billion & [2.4 billion; 4.4 billion] & 4.8 & [3.5;6.2]\\ \hline
    High growth & 4.0 billion & [2.9 billion; 5.1 billion] & 5.7 & [4.1;7.3]\\ \hline
    Extreme growth & 4.8 billion & [3.5 billion; 6.2 billion] & 6.8 & [4.9;8.9] \\ \hline
    \end{tabular}
    \caption{Valuation and share value of Zynga in the base case, high growth and extreme growth scenarios.}
    \label{table}
\end{table}

\noindent We obtain a valuation of 3.4 billion USD for our base case scenario, well below the $\approx$ 7 billion USD value at IPO or the 9 billion value at the end of March, 2012. Even the unlikely extreme growth case scenario could not justify any of the valuations we have seen in the market so far.

\section{Historic evolution} \label{comparison}

\begin{figure}[!h] 
    \centering
    \includegraphics[width=1\textwidth]{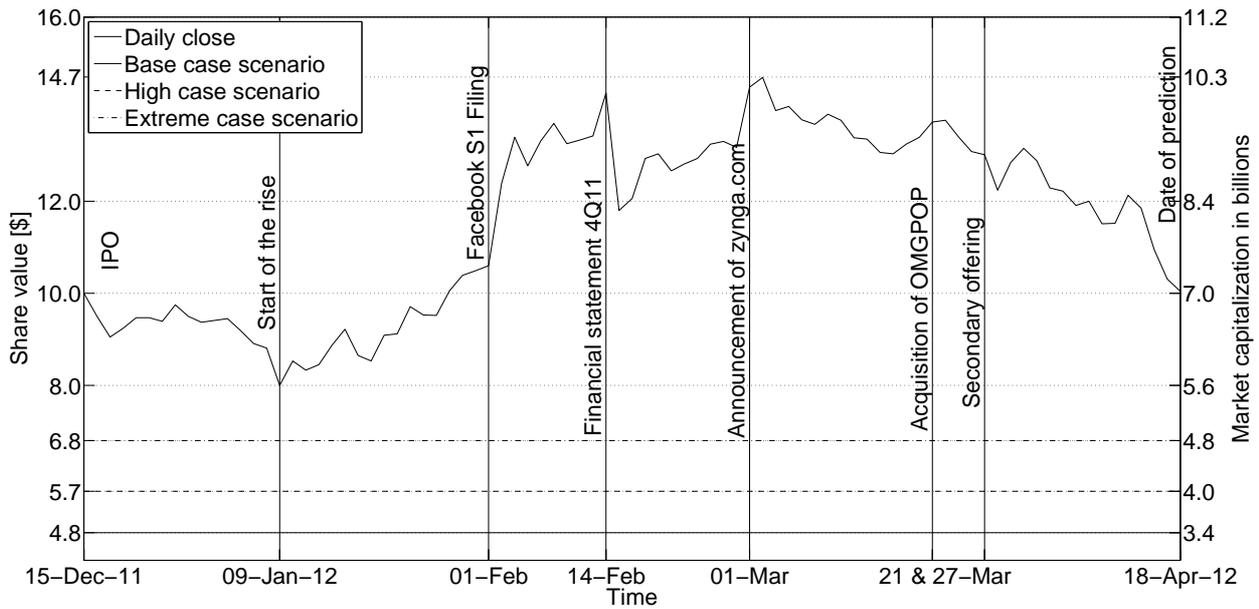}
    \caption{Share value of Zynga over time. The base case, high and extreme growth scenarios are represented (horizontal lines) as well as important dates for the stock price (vertical black lines). (Source of the data: \cite{YahWeb})}
    \label{share_value2}
\end{figure}

At the time of the IPO, on December 15\textsuperscript{th}, 2011, Zynga was valued at 7 billion USD. Right after the IPO, on December 27\textsuperscript{th}, we published an article on Arxiv \citep{Foretal11} pointing to an overvaluation of Zynga (which was estimated at 4.2 billion USD in our base case scenario). Since then, Zynga published its earnings for the 4th quarter of 2011. These figures increased the accuracy of our valuation since they contributed to reduce the difference between the three scenarios for the revenues per user. By now, it has traded on the stock market for almost 3 months. The big question is whether the share price of Zynga moved into the direction of its fundamental value. As we can see in figure \ref{share_value2}, it was quite the contrary: after an initial depreciation of the share value reaching a minimum of 7.97 USD on January 9\textsuperscript{th} (still above our extreme case scenario), it was followed by a moderate run-up in price until February 1\textsuperscript{st} 2012, the date of the S1 filing from Facebook. After that, without any solid economic justification, Zynga skyrocketed to a maximum of 14.55 USD/share corresponding to a 10.2 billion USD valuation on February 14\textsuperscript{th}. However, after the release of the 4th quarter results, the company lost more than $15\%$ in a single day, regaining a part of this loss on the following days and peaking again at 14.62 USD on March 2\textsuperscript{nd} , 2012. On March 28\textsuperscript{th}, insiders of Zynga (including its CEO, Mark Pincus) sold 43 million shares in a secondary offering (see section \ref{timeline}) over the counter for 12 USD/share (\citet{SecWeb4}). Zynga's shares subsequently experienced a 6\% drop in one day. By the end of April 18\textsuperscript{th}, the publication date of our trading strategy (see section \ref{arbitraging}), Zynga closed at 10 USD. More details about Zynga's price trajectory are given in table \ref{price_trajectory} and figure \ref{share_value2}.

\begin{table} [!h]
    \centering
    	\begin{tabular}{ | l | r| p{10cm} |}
    	\hline
    	Date & Share price [\$] &Event  \\ \hline
    	2011-12-15 & 10.00 & Zynga goes through its IPO. \\ \hline
	2012-01-09 & 8.00 & Zynga closes at its lowest level for the next 4 months. \\ \hline
	2012-02-01 & 10.60 & Facebook publishes its S1 Filing. This fuels Zynga's bubble. \\ \hline
	2012-02-14 & 14.35 & Zynga unveils its financial results for the 4\textsuperscript{th} quarter of 2011. The next day, Zynga's share price experiences an 18\% drop, its biggest drop until today. \\ \hline 
	2012-03-01 & 14.48 & Zynga announces that it will launch zynga.com (\citet{ZynCom}), an independent platform from Facebook. The news is followed by a small increase in share price. \\ \hline
	2012-03-21 & 13.72 & Zynga acquires OMGPOP, another social gaming company, for over 200 million \$ (\citet{OMG}). \\ \hline
	2012-03-27 & 13.01 & Inside investors of Zynga (including its CEO, Mark Pincus) sell 43 million shares at 12\$/piece in a secondary offering. This is followed by a significant 6\% drop in share price. \\ \hline
	2012-04-18 & 10.04 & We publish our short-term prediction on Arxiv. \\ \hline
    	\end{tabular}
    \caption{Important events in Zynga's price history until April 18\textsuperscript{th}, 2012, the date of our prediction.}
    \label{price_trajectory}
\end{table}

\begin{figure}[!h] 
    \centering
    \includegraphics[width=1\textwidth]{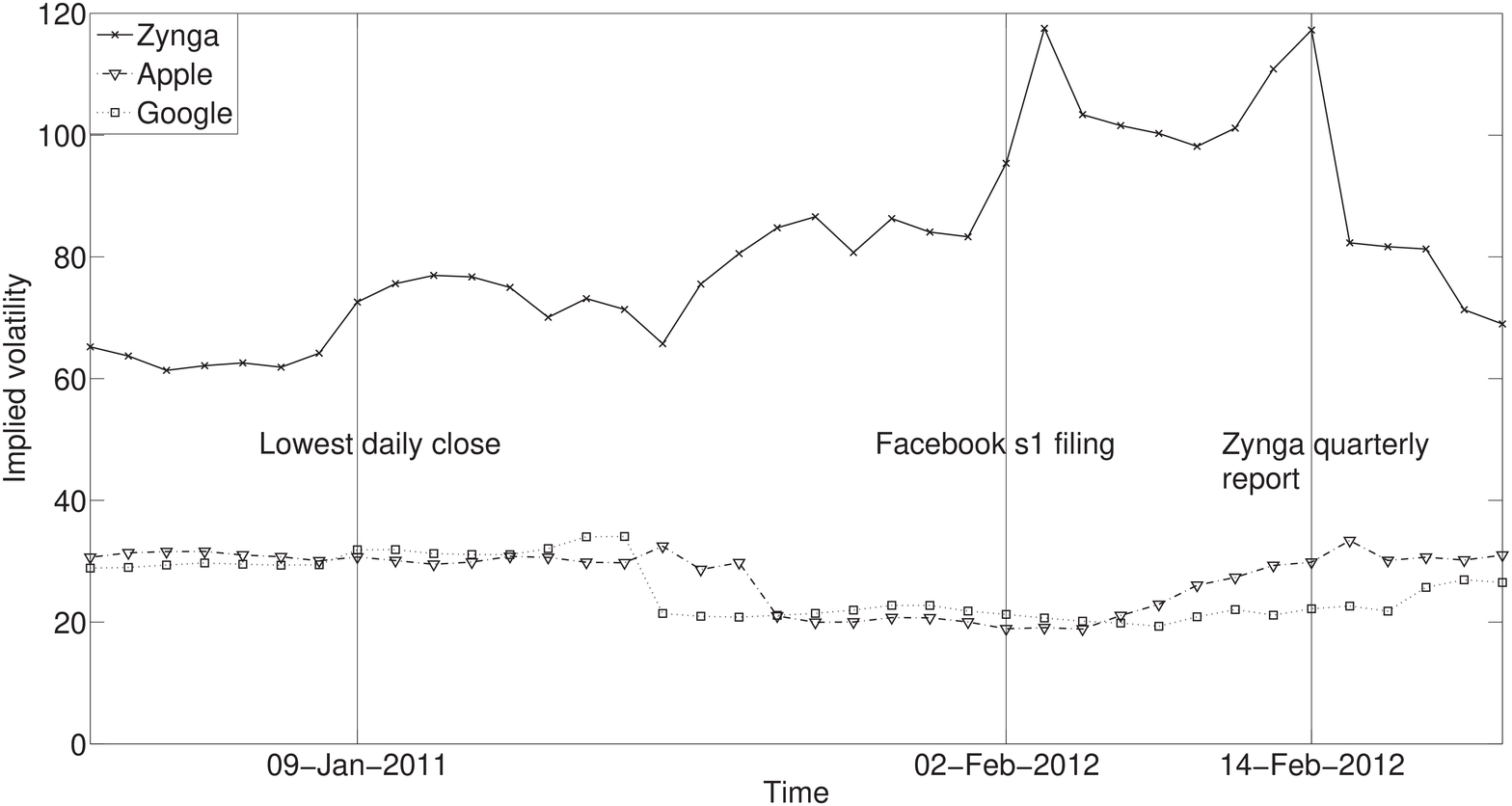}
    \caption{Implied volatility of Zynga, Apple and Google. Zynga has a much higher and less stable volatility than Apple or Google. (Source of the data: Bloomberg)}
    \label{implied_vol}
\end{figure}

\noindent The highly volatile, news driven behavior of Zynga's stock price can be quantified using the implied volatility measure. In option pricing, the value of an option depends among others on the volatility of the underlying asset. Knowing the price at which an option is traded, one can reverse engineer the implied volatility, the volatility needed to obtain the market value of the option given a pricing model. This standard measure has the upside of being forward-looking: contrary to the historic volatility, the implied volatility is not computed from past known returns. As such, implied volatility is a good proxy for the mindset of the market. Figure \ref{implied_vol} compares the implied volatility priced by the market for options written on Zynga with that of Google and Apple.\\

\noindent We can observe a big difference between the two groups. While Apple and Google have a standard stable implied volatility, there is much more uncertainty surrounding Zynga in the eyes of the market, given its high and unstable volatility. What this tells us, until now, is that the market players have a hard time putting a value on Zynga. This perception is reinforced by the following event: on February 15\textsuperscript{th}, 2012, the day following the biggest drop of Zynga since its IPO, most investment banks (with some exceptions) downgraded Zynga's stock rating, readjusting their price target. Some notable examples of actual price targets per share are \citep{BesWeb}:
\begin{itemize}
\item Barclays Capital: $11\$$
\item BMO Capital Markets $10\$$
\item Evercore Partners: $10\$$
\item JP Morgan Chase: $15\$$
\item Merrill Lynch: $13.5\$$ 
\item Sterne Agee $7\$$
\end{itemize}

\noindent Compared with our analysis (see table \ref{table}), these price targets seem to be high. Moreover, even among ``experts", the differences can be significant (the price target of Sterne Agee is less than half of JP Morgan Chase's). One should however keep in mind that the recommendations of most of these companies may not be independent of their own interest, as for example the fact that JP Morgan, Merril Lynch and Barclays Capital are underwriters of the IPO (see \cite{MicWom99, Decetal00}).\\ 

\noindent We will have to wait and see how Zynga evolves on longer time scale, but so far, our analysis indicates that Zynga is in a bubble, its price not being reflected by its economic fundamentals. Zynga may be the symptom of a greater bubble, affecting the social networking companies in general, as suggested by the overpricing of Facebook and Groupon \citep{CauSor12}.

\section{Arbitraging Zynga's bubble}  \label{arbitraging}

While the market price of Zynga should converge to its fundamental value on the long run, a prediction of its price movements on a shorter time scale is difficult. It is however possible 
to develop investment strategies by 
using a combination of our determination of Zynga's intrinsic value done above with a well known phenomenon, namely the drop of market price when insiders are allowed to sell their shares 
at the end of their lock-up period.

\subsection{End of lock-up and its implication} \label{prediction}
When a company goes public, only a fraction of their shares are put on the market (14\% in the case of Zynga). The rest of the shares are locked-up for a period of typically 180 days. It is common for IPOs to have a lock-up period in order to prevent insiders from massively selling their shares after the IPO, and thus driving the market value of the company down. There is a vast amount of literature exploring the effect of the end of the lock-up period on the share value of a company. While there is a broad consensus on the fact that companies, on average, experience abnormal negative returns following the end of the lock-up period, different authors give different explanation of this effect. %Some of the relevant results for Zynga are the following: 
\cite{FieHan01} find that venture capital backed firms experience the largest price drop at the end of the lock-up period. \cite{Braetal01} confirm the finding and add that the ``quality" of the IPO underwriters as well as the price increase since the IPO is positively correlated with the drop in share value. \cite{Gao05} finds that firms with the highest forecast bias and the highest forecast dispersion by analysts experience the largest drop. Finally, \cite{Ofe08} makes the argument that a significant increase in share supply can explain the price drop subsequent to the end of the lock-up period. He further argues that the higher the stock price volatility before the end of the lock-up, the bigger the drop in share value. 

While most of the above mentioned authors find that it is difficult to develop an arbitrage strategy to take advantage of this effect, we should stress that they all based their works on samples of companies independent of any view regarding their intrisinc value. If they could bias their sample towards the companies whose market value is significantly higher than their fundamental value, we would expect a different outcome. We hypothesize that the overvaluation of the company would be reflected in its market price, as soon as insiders, better informed about the fundamentals of their company, would be allowed to trade freely, i.e., at the end of the lock-up period. We believe that the information asymmetry between outside traders (the only ones who are allowed to trade the shares of Zynga from its IPO) and insiders would be incorporated in the price formation of such a company, and move its market price towards its fundamental value.

\subsection{Prediction for Zynga} \label{secondary}

\subsubsection{Timeline} \label{timeline}
On March 23\textsuperscript{rd}, Zynga announced, in an \citet{SecWeb3}, that inside investors including CEO Mark Pincus would sell about 43 million shares. This move was surprising, since Zynga's inside investors were subjected to a lock-up period ending on May 29\textsuperscript{th}. However, a secondary offering was authorized by the underwriters of the IPO and was concluded on March 28\textsuperscript{th} with the shares being sold over the counter for \text{$12\$$/piece} (0.36$\$$ of which went to the underwriters). On that day, the share value of Zynga experienced a large drop, going from 13.02 to 12.24 $\$$ (this corresponds to a 6\% decrease). It should be noted that, subsequent to this transaction, these insiders are again subject to a lock-up period and won't be able to trade until its end. In practice, the remaining locked shares will be released in the market in several steps (see \citet{SecWeb4}):
\begin{enumerate}
\item Approximately 115 million shares held by non-executive employees around April 30\textsuperscript{th} (or 3 days after Zynga will disclose its financial statement for the first quarter of 2012).
\item Approximately 325 million shares held by non-employee stockholders that have not participated in the secondary offering (see section \ref{secondary}) on May 29\textsuperscript{th}, 2012.
\item Approximately 50 million shares held by directors, executive employees and the stock holders who participated in the secondary offering (such as Mark Pincus) on July 6\textsuperscript{th}, 2012. \label{pincus}
\item Approximately 150 million shares held by the same persons as in \ref{pincus}. on August 16\textsuperscript{th}, 2012.
\end{enumerate}

\subsubsection{Effect of the financial results for the first quarter of 2012}

\begin{figure}[!h] 
    \centering
    \includegraphics[width=1\textwidth]{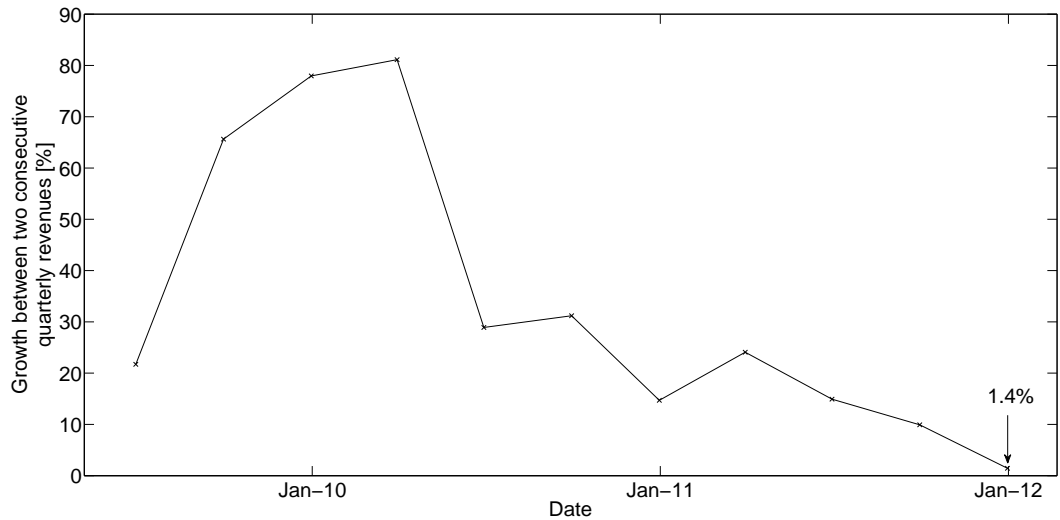}
    \caption{Percentage difference between the revenues of two consecutive quarters. We can see that Zynga's performance in the last quarter was very poor. This suggests that taking the 1.4\% figure as a benchmark to evaluate Zynga's performance for the next quarter may lead investors to be overly optimistic, especially during a bubble period (Source of the data: S1 Filing to the SEC).}
    \label{quarterly_growth}
\end{figure}

When trying to predict the future price movements of Zynga, one cannot ignore the fact that on April 26\textsuperscript{th}, 3 days before the end of the first part of the lock-up period, the company will release its financial results for the first quarter of 2012. To understand the implications of this report on Zynga's price movement, we base ourselves on our diagnostic of a bubble. Indeed, during a bubble, phenomena like herding and imitation are dominant among traders (\citealp{Sor03}). As such, the market players are very sensitive to new information, giving rise to behaviors inconsistent with its content. We believe that the release of Zynga's financial statement on April 26\textsuperscript{th} can be such an event. According to our model, Zynga's yearly revenue per user is saturating. This can be seen on the left-hand side of figure \ref{rev_per_dau}, where the yearly revenue per user, computed at each quarter as the running sum of the four previous quarters, are well fitted by a logistic function. The saturation of Zynga's yearly revenues per user is a powerful argument 
for the diagnostic of a bubble in 
Zynga's market valuation. Even with a hypothetical 357 million USD of revenues for the first quarter of 2012 (which will be published on April 26\textsuperscript{th}) meaning a 15 \% increase from the 311 million USD of revenues last quarter, the yearly revenues per user would fall right onto our logistic fit. Hence, even a 15 \% increase in quarterly revenues would not be sufficient to rationally reject the saturating trend of Zynga's revenues per user that we predict. However, compared with the results of the previous quarter, which saw Zynga's revenue only rise by 1.4\% (see figure \ref{quarterly_growth}), this would be seen as a very strong performance and would most likely be followed by an increase in share value, even more so in the bubble environment that we diagnose. 

\subsubsection{Effect of the end of the lock-up period}

Up to now, there are about 150 million shares tradable on the market (100 million from the IPO and about 50 million from the secondary offering). The 115 million shares coming to the market around April 30\textsuperscript{th} represent an important increase in the free-floating shares of Zynga. As such, and because Zynga satisfies most of the conditions given in subsection \ref{prediction} leading to a large price decrease, we predict a drop of Zynga's market value around that date. We should mention that what happens around April 30\textsuperscript{th} will be conditional on what will have happened on April 26\textsuperscript{th}: we predict this drop to be larger if Zynga's stock price increases on April 26\textsuperscript{th} and smaller if the stock price decreases on the same date. Such a phenomenon could take place at each such dates.

\subsubsection{Proposed strategy}

We believe that there is a high probability for strong corrections in Zynga's price after each partial lift of the lock-up period. While we have shown that even an apparently strong performance on April 26\textsuperscript{th} would be in line with our diagnostic of Zynga's saturating revenues per user, one should not be surprised to see its share value rise in this bubble environment. On the long-run, we predict Zynga's market value to converge to its intrinsic value of 3.4 billion USD.

In summary, the proposed strategy is based on three time periods:
\begin{enumerate}
\item From the time of writing (April 16\textsuperscript{th}, 2012) to the announcement of the financial results (around April 26\textsuperscript{th}, 2012): stay out of Zynga or hedge if invested.
\item From the day after the earnings announcement (around April 27\textsuperscript{th}, 2012) to the end of the first lock-up period (around April 30\textsuperscript{th}, 2012): if the financial results are significantly 
above those of the previous quarter, buy Zynga for a short term holding period. Otherwise short it.
\item From the end of the first lock-up period (after April 30\textsuperscript{th}, 2012): close all open long positions and short. Monitor the subsequent quarterly releases and the successive ends of future lock-up periods to 
position a strategy in the same spirit as above.
\end{enumerate}

\section{Conclusion} \label{conclusion}

\noindent In this paper, we have proposed a new valuation methodology to price Zynga. Our first major result is to model the future evolution of Zynga's DAU using a semi-bootstrap approach that combines the empirical data (for the available time span) with a functional form for the decay process (for the future time span). \\

\noindent The second major result is that the evolution of the revenues per user in time, $r_i$, shows a slowing of the growth rate, which we modeled with a logistic function. This makes intuitive sense as these $r_i$ should be bounded due to various constraints, the hard constraint being the economic one, since Zynga's players only have a finite wealth. We studied 3 different cases for this upper bound: the most probable one (the base case scenario), an optimistic one (the high growth scenario) and an extremely optimistic one (the extreme growth scenario).\\

\noindent Combining these hard data and soft data revealed a company value in the range of 3.4 billion to 4.8 billion USD (base case and extreme growth scenarios).\\

\noindent On the basis of this result, we can claim with confidence that, at its IPO and ever since, Zynga has been overvalued. Indeed, even the extreme growth scenario (implying 43 USD/DAU at saturation) would not be able to justify any value the company had until now. It is worth mentioning that we adopted a rather optimistic approach:
\begin{itemize}
\item We have taken a (slow) power law for the decay process (even in cases where exponentials might be better).
\item We chose games only in the top 20 with equal probability in the simulation process (implying that there is the same probability to create a top game and an average/unsuccessful one).
\item We took a $15 \%$ profit margin and supposed that all the future profits would be distributed to the shareholders.
\item We implicitly assumed that the real interest rates and the equity risk premium stay constant at $0\%$ and $5\%$ respectively, for the next 20 years \citep{CauSor12}.
\end{itemize}

\noindent Given these optimistic assumptions, all our estimates should be regarded as an upper bound in our valuation of Zynga. \\

We should also stress that our assumptions do take into account the innovations that Zynga will have to create in order
to continue its business, akin to the Red Queen's race in Lewis Carroll' novel with Alice constantly running but remaining in the same spot.\\

While the fundamental value computed here suggests that Zynga's market price should decrease on the long term, we were able to delineate an investment strategy built on the expected future price movements of Zynga on a short term scale. In particular, we believe that Zynga's share value will drop significantly around April 30\textsuperscript{th}, 2012 after a possible short run-up
following the announcement of the financial results of the first quarter around April 26\textsuperscript{th}, 2012.

\section{Post-mortem analysis of the proposed strategy (added on May 24\textsuperscript{th}, 2012)}

\begin{figure}[!h] 
    \centering
    \includegraphics[width=1\textwidth]{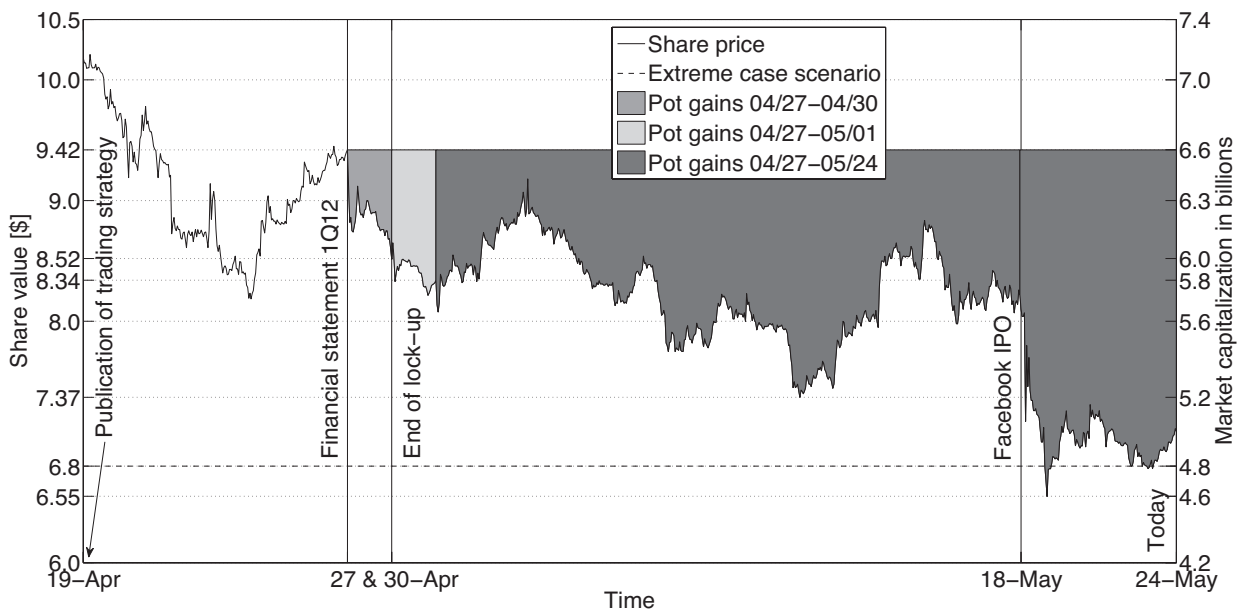}
    \caption{Price dynamics of Zynga from the publication of our trading strategy on April 19\textsuperscript{th}, 2012 on the arXive until May 24\textsuperscript{th}, 2012 just before going to press. Potential gains (indicated as ``Pot gains'' in the inset) 
    that could be obtained by opening a short position on April 27\textsuperscript{th} are indicated by the shaded area. The data has a 10 minutes time resolution. (Source of the data: Bloomberg).}
    \label{post_mortem}
\end{figure}

This section was added on May 24\textsuperscript{th} after the main paper was accepted for publication on May 17\textsuperscript{th}, 2012. The version of the paper with our ex-ante proposed strategy (section \ref{arbitraging}) can be found on Arxiv with the date stamp of April 19\textsuperscript{th}, 2012 \citep{Foretal12}. In this section, we will evaluate our ex-ante prediction in the light of the most recent events. Figure \ref{post_mortem} summarizes the price movements of Zynga from April 19\textsuperscript{th} till May 24\textsuperscript{th}, 2012. The strategy was based on 3 legs:

\begin{enumerate}
	\item
	\begin{it}
	From the time of writing (April 16\textsuperscript{th}, 2012) to the announcement of the financial results (around April 26\textsuperscript{th}, 2012): stay out of Zynga or hedge if invested.\\
	\end{it}
	Between April 19\textsuperscript{th} and April 26\textsuperscript{th}, Zynga's share price dropped from 10.2 to 8.2 USD and then rebounced to 9.42 USD (the opening price on April 19\textsuperscript{th}). 		Although the stock went down 7.7\% in a 	week, its behavior was very volatile. As we did not have any strong factual information to support a clear trading strategy before April 								27\textsuperscript{th}, not taking a position appears to have been an acceptable advice.

	\begin{it}
	\item From the day after the earnings announcement (around April 27\textsuperscript{th}, 2012) to the end of the first lock-up period (around April 30\textsuperscript{th}, 2012): if the financial results are 			significantly 
	above those of the previous quarter, buy Zynga for a short term holding period. Otherwise short it.\\
	\end{it}
	This part was undeniably a success. On April 26\textsuperscript{th}, after the markets closed, Zynga revealed its financial results for the first quarter of 2012. Its quarterly revenues were weak, since they only 		grew 3.1\% since the previous quarter, confirming that the company is in its saturation phase. As a result, on April 27\textsuperscript{th}, Zynga experienced a drop of 9.6\%, one of its largest daily drops since its 	IPO.

	\begin{it}
	\item From the end of the first lock-up period (after April 30\textsuperscript{th}, 2012): close all open long positions and short. Monitor the subsequent quarterly releases and the successive ends of future 	lock-		up periods to position a strategy in the same spirit as above.\\
	\end{it}
	As this last part covers a large time-period (from April 30\textsuperscript{th} to May 24\textsuperscript{th}, 2012), we divide it into a short-term part (the first day) and a longer-term part (until the time of writing of the post-mortem analysis, 
	May 24\textsuperscript{th}, 2012). 
		\begin{itemize}
		\item On the first day (April 30\textsuperscript{th}), the prediction was proven successful. Indeed, as a result of the end of the lock-up period, the stock further dropped 2.1\% in a single day. Had someone 			opened a short position on April 27\textsuperscript{th} (at the opening of the markets) and closed it on April 30\textsuperscript{th} (at the closing of the markets), he would have benefited of a 11.5\% drop 			over 2 trading days.
		\item On the longer term, the price trajectory, although quite volatile, went significantly down. This was accentuated by Facebook's IPO on May 18\textsuperscript{th}. Indeed, it was soon clear from the price 		dynamics after the Facebook's IPO that it was not a big success. On the other hand, due to the use of the ``over-allotment" or ``green shoe" option, the price would be kept artificially above the IPO price for 		a time. Therefore, investors targeted other social networks like Zynga which lost 13\%, Linkedin which lost 6\%, Groupon which lost 7\% or Renren, the Chinese Facebook which lost 21\%. The lack of 				rebound of Zynga (until today) may be due to the loss of its status as a ``proxy" for Facebook. It is worth noting
that, for the first time since it went public, Zynga's value entered our fundamental valuation 		bracket, when on May 21\textsuperscript{st} it dropped to 6.5\$/share (intraday), below our extreme case scenario of 6.8\$/share... The future will tell us if this price dynamic will stabilize close to our 				fundamental value calculation or if the investors' realistic perception of Zynga's is only temporary.
		\end{itemize}
\end{enumerate}

\noindent To sum up, we have successfully predicted the downward trend of Zynga. Since April 27\textsuperscript{th}, 2012, Zynga dropped 25\%.

\subsection*{Acknowledgments}
\noindent The authors would like to thank Ryohei Hisano, Vladimir Filimonov and Susanne von der Becke for useful discussions.

%\newpage
%\section{Appendix}
%A verification of the Poisson hypothesis was done as follows: for the innovation process, we drew $\Delta t$ from its empirical distribution, instead of sampling $\Delta t$ from its exponential counterpart $p(\Delta t) \propto e^{-\lambda \Delta t}$ (consequence of having a Poisson process). This bootstrapping method was justified by the independence of the waiting times (section \ref{independence}). The idea behind the test was to see whether the valuation obtained from this method was in good agreement with the valuation obtained with the Poisson hypothesis (\ref{valuation}). Results are shown in table \ref{poisson_vs_bootstrap}. 
%
%\begin{table} [!h]
%    \centering
%    \begin{tabular}{ | l | c | c | c |}
%    \hline
%    Method & Market cap & $95\%$ two-sided confidence interval & relative difference \\ \hline
%    Poisson & 3.36 billion & [2.43 billion; 4.35 billion] & \multirow{2}{*}{$1.2\%$/[$3.3\%$;$0.7\%$]}  \\ \cline{1-3}
%    Bootstrap & 3.40 billion & [2.51 billion; 4.38 billion] &  \\ \hline
%    \end{tabular}
%    \caption{Valuation of Zynga using a Poisson process and a bootstrap method for the innovation process.}
%    \label{poisson_vs_bootstrap}
%\end{table}
%
%\noindent It can be seen, that both methods result in very similar valuation for Zynga confirming that the Poisson process is a valid model for its innovation.

\bibliographystyle{abbrvnat}

\bibliography{zynga_valuation_short2}

\end{document}